\documentclass[nofootinbib]{revtex4}
\usepackage[T1]{fontenc}
\usepackage{amsmath,amssymb}
\usepackage{epsfig}
\usepackage{dcolumn}
\usepackage{graphicx}
\usepackage[usenames,dvipsnames]{color}
\usepackage{slashed}
\usepackage[colorlinks,citecolor=blue]{hyperref}
\usepackage{pdfpages}
\usepackage{float}
\usepackage{adjustbox}
\usepackage[autostyle]{csquotes}
\usepackage{subcaption}
\begin{document}

\title{\boldmath Exploring dynamics of $A_4$ flavour symmetry using low scale seesaw mechanisms}
\author{ Kalpana Bora$^{1}$ \footnote{kalpana@gauhati.ac.in}} 
\author{ Maibam Ricky Devi$^{1}$ \footnote{deviricky@gmail.com}} 
\affiliation{Department of Physics, Gauhati University, Assam, India$^{1}$}

\begin{abstract}
Low scale seesaw models, like low scale type II, inverse (ISS),  and linear seesaw (LSS) models provide an interesting mechanism to obtain light neutrino masses and mixings, as they can be tested in future TeV scale experiments. Discrete flavour symmetry groups like $A_4$ can be incorporated to explain the flavour structure of particles. But, so far, nothing is known about dynamics of flavour symmetry - scale of its breaking, or VEV alignment of the flavon fields. In a recent study \cite{Ricky}, we have investigated and shed light on how to pinpoint the favoured VEV alignment of the flavon field using light neutrino oscillation data. In this work, for the first time, we present an analysis on dependence of light neutrino masses on scale of flavon VEV in these three seesaw models, and comment on how this information can be used to discriminate among them. We also discuss about the estimated value of the constant $F$ which can constrain various coupling constants of the model, cut-off scale of the theory and scale of flavour symmetry breaking. This study can provide useful insight into the hitherto unknown dynamics of flavour symmetry and hence can contribute as an important ingredient in the model building for future studies.

\end{abstract}
\maketitle

\section{\label{sec:level1}Introduction}

Fundamental particles are found to follow certain pattern among them, with reference to their masses and coupling strengths among themselves. Since in standard model (SM), there are three families (generations) of particles, the flavour group that can be used to describe them must have a triplet in it. Many flavour groups have been used to explain this flavour structure of particles - $A_4$, $S_3$, $S_4$ to name a few. Among these, $A_4$ is the simplest group that can accommodate all the particles of the SM, and hence in this work we have chosen this group for our analysis. Flavour symmetry is not observed at low scales and hence it must be broken, and this can be done by desiring that the flavon fields acquire VEV (vacuum expectation value).  Flavon fields couple to the particles of the theory to give desired coupling or masses to them. Though many scenarios have been proposed, but there are no experimental proofs in support of scale of this flavour symmetry breaking or the direction in which the vacuum stabilizes.\\

With this motivation, recently we proposed a scenario \cite{Ricky} in which the favoured direction of the VEV of these flavon fields can be pinpointed in low scale seesaw models with $A_4$  flavour symmetry, with the help of some other discrete symmetries like $Z_3$, $Z_4$ etc. We constructed the Lagrangian such that the light neutrino mass matrix produces the neutrino oscillation parameters within the $3\sigma$ range of their global best fit values. These models have the advantage that they can be tested in near future experiments, as the seesaw scales involved are of the order of few TeV. We chose a benchmark value of $\dfrac{(\textrm{flavon VEV})}{\Lambda}$=0.1, for three low scale seesaw models, where $\Lambda$=100 TeV is the cut-off scale of the theory. So, the scale of the flavour symmetry breaking was considered to be $\sim 10$ TeV, and we successfully obtained the light neutrino masses and mixing pattern lying within the range of their global best fit values.\\

In this work, extending the work of our earlier work \cite{Ricky}, we present an interesting analysis on dynamics of the flavour symmetry breaking, i.e., how the light neutrino mass depend on the ratio $\dfrac{(\textrm{flavon VEV})}{\Lambda}$ in the three seesaw models. If in future experiments, we are able to obtain this dependence of light neutrino mass on flavon VEV, then through this study one can discriminate among the models. It may be noted that so far, no information is available either on favoured VEV alignment of the flavon VEV, or on the scale of flavour symmetry breaking,  and hence the findings of this work are important and may contribute in model building with flavour symmetries. For  some of the $A_4$ symmetry related works, the interested reader can refer to  \cite{Dinh:2016tu}-\cite{Cai:2018upp}.

\section{Dynamics of flavour symmetry}

In \cite{Ricky}, we discussed in detail, the  structure of the Lagrangians in low scale type-II seesaw \cite{Mohapatra:1980yp}, ISS \cite{PhysRevLett.56.561}-\cite{PhysRevD.34.1642},  and LSS \cite{Malinsky:2005bi}-\cite{Deppisch:2015cua} models with $A_4$ flavour symmetry. We first constructed the Lagrangians in three respective models, so that no unwanted terms appear in it, then obtained the light neutrino mass matrices in all of them. Finally, we compared them with the phenomenological neutrino mass matrix and obtained the equations for unknown flavon VEVs, solving them using computation we obtained correlation plots among the unknown neutrino oscillation parameters like - light neutrino mass, atmospheric mixing angle $\theta_{23}$ (as its Octant is not yet fixed), and leptonic CP violating phases. For more details, interested reader may refer to \cite{Ricky}. There we found that the light neutrino mass can be expressed as:

\begin{equation}
m_{\nu}= F\times (\textrm{flavon VEV}).
\end{equation}\\
The expressions of the dimensionless coupling constant F for  the three seesaw models can be summarized as follows \cite{Ricky}:\\
\begin{eqnarray}
 F=\dfrac{y\upsilon_{\Delta_{L}}}{\Lambda}, \textrm{for type-II seesaw model,} & &
\label{F:typeII}
\end{eqnarray}

\begin{eqnarray}
 F= \dfrac{Y^{2}_{D}Y_{\mu}}{Y^{2}_{M}}[\dfrac{\upsilon^{2}_{h}\upsilon_{\rho^{\prime}}\upsilon^{\dagger}_{\rho^{\prime \prime}}}{\Lambda^{4}}], \textrm{for ISS model,} & &
\label{F:ISS}
\end{eqnarray}
\\
\begin{eqnarray}
 F=\dfrac{2 Y_{L}Y_{D}\upsilon^{2}_{h}\upsilon^{\dagger}_{\varepsilon}}{\Lambda^{2}Y_{M}\upsilon_{\rho}} , \textrm{for LSS model.} & &
\label{F:LSS}
\end{eqnarray}
\\
Here, $\upsilon_{\Delta_L}$ is the VEV of the additional (beyond SM) SU(2) triplet scalar field that is responsible for generating light neutrino mass in type-II see saw model, $y_i$($Y_i$) are dimensionless coupling constants of various flavon fields with other fields of the SM in low scale type-II seesaw(ISS, LSS) models, and $\upsilon_{\rho}$ etc are the VEVs of various flavon fields. In ref.[1],  we had used  same values of $ \dfrac{\langle flavon \rangle}{\Lambda}$, lightest neutrino mass and heavy mass scale, in the three models, just for the sake of comparison among them. The value of constant F, as computed while solving simultaneous equations, was found to be  $\sim 10^{-2}$. Now the value of light neutrino mass scale can be fine-tuned (to suit the experimentally observed values), if we adjust the values of $ \dfrac{\langle flavon \rangle}{\Lambda} $ and the seesaw scale, since we don't know any of them. Hence the couplings of the theory, and these two scales can be adjusted, with the constraint that\\
\begin{equation}
F\sim 10^{-2} \textrm{ (dimensionless constant)}
\end{equation}\\
as is evident from Equations (\ref{F:typeII}),(\ref{F:ISS}), and (\ref{F:LSS}). For example, as mentioned earlier, in say ISS,
\begin{equation}
 \dfrac{Y^{2}_{D}Y_{\mu}}{Y^{2}_{M}}=\Lambda  \times 1.52\times 10^{-21} \textrm{eV}^{-1}
\end{equation} \\
for  $ \dfrac{\langle flavon \rangle}{\Lambda} =10^{-1}$, But same F can also be obtained for  $ \dfrac{\langle flavon \rangle}{\Lambda} = 10^{-2}$, and then the couplings would satisfy \\
 \begin{equation}
 \dfrac{Y^{2}_{D} Y_{\mu} }{Y^{2}_{M}}=\Lambda  \times 1.52\times 10^{-18} \textrm{eV} ^{-1}
\end{equation} \\
Another interesting feature observed from Equation  (\ref{F:typeII}),(\ref{F:ISS}), and (\ref{F:LSS}) is the dependence of light neutrino mass, $ m_{\nu} $ on 
$ \left(  \dfrac{\langle flavon \rangle}{\Lambda}\right)   ^{x}$, where $ x=1 $ in type II and LSS, and  $ x=3 $ in ISS model.\\
\begin{equation}
m_{\nu} \propto \left(  \dfrac{\langle flavon \rangle}{\Lambda}\right)   ^{x}
\label{eqn:58}
\end{equation} \\
\begin{table*}
\centering
\begin{adjustbox}{width=16 cm}
\begin{tabular}{|c|c|c|c|c|}
\hline 
Seesaw model & $  m_{\nu}$ (eV) & for $ \Lambda=100 $ TeV, $ \dfrac{\langle flavon \rangle}{\Lambda}=0.1  $ & $ \Lambda=1000 $ TeV, $ \dfrac{\langle flavon \rangle}{\Lambda}=0.1  $ \\ 
\hline 
Low-scale type-II seesaw  & $ y\upsilon_{\Delta}\dfrac{flavon}{\Lambda} $  &  y=0.1 and $ \upsilon_{\Delta}=10 $ eV &  y=0.1 and $ \upsilon_{\Delta}=10 $ eV \\ 
\hline 
Inverse seesaw  & $ \dfrac{Y^{2}_{D} Y_{\mu}}{Y^{2}_{M}}(\upsilon^{2}_{h} \upsilon_{\rho^{\prime}}\upsilon^{\dagger}_{\rho^{\prime \prime}})\dfrac{ flavon}{\Lambda^{4}} $ & $ \dfrac{Y^{2}_{D}Y_{\mu}}{Y^{2}_{M}} =1.52 \times 10^{-7}$  & $ \dfrac{Y^{2}_{D}Y_{\mu}}{Y^{2}_{M}} =1.52 \times 10^{-6}$   \\ 
\hline
Linear seesaw  & $ \dfrac{2Y_{L}Y_{D}\upsilon_{h}^{2}\upsilon_{\epsilon}^{\dagger}}{Y_{M}\upsilon_{\rho}}\dfrac{ flavon}{\Lambda^{2}} $ & $ \dfrac{Y_{L}Y_{D}}{Y_{M}}= 7.63\times 10^{-10} $  &  $ \dfrac{Y_{L}Y_{D}}{Y_{M}}= 7.63\times 10^{-9} $ \\ 
\hline

\end{tabular}
\end{adjustbox}
\caption{Values of light neutrino mass $m_\nu$ for some chosen values of couplings and flavon VEVs for three  low-scale seesaw models. It shows the correlations among them, and that how $m_\nu$ scales with flavon VEV. Here, $\upsilon_{\rho}$, $\upsilon_{\epsilon}$ etc are VEVs of various flavon fields \cite{Ricky}}.
\label{tab:Yscales}
\end{table*}\\
In Table 1, we show the implications of Eq. (8), for the purpose of demonstration - for similar decrease in light neutrino mass $m_{\nu} $ (say $ \dfrac{m_{\nu}}{y} $), keeping coupling constant, the $\dfrac{\langle flavon \rangle}{\Lambda}$ decreases by a factor of $y$, in type II and LSS, while in  ISS, it decreases by a factor of $\dfrac{1}{y^{1/3}}$. This implies that if experimentally observed value of light neutrino mass decreases by a certain factor y, the scale of flavour symmetry decreases less in ISS ($ \dfrac{1}{y^{1/3}} $). \\

Eqs. (6-8) and Table 1 are the novel results of our analysis, that demonstrates new correlations among scale of symmetry and various coupling constants, and are testable in future measurements. They give us useful insight into the dynamics of flavour symmetry - that how the scale of flavour symmetry breaking and various coupling constants can be adjusted, and hence pinpointed so as to match these predictions with experimental measurements.

\section{Summary}
\label{sec:4}
To summarise, in this work, we presented a novel analysis on the dynamics of $A_4$ flavour symmetry and new correlations among scale of symmetry and various coupling constants, in context of low scale type-II, inverse seesaw (ISS) and linear seesaw (LSS) models. Our new results  are summarised in Eqs. (6-8) and Table 1. We found interesting correlation among light neutrino mass, scale of flavour symmetry breaking and various coupling constants of the theory, which can be tested in future measurements. Thus, this study can provide insightful guidelines on the important issue of dynamics of flavour symmetry, which is still not resolved in flavour physics  (in particular the VEV scale of flavon and their coupling constants to other particles), and we hope that future experiments would be able to support or refute the ideas presented here.\\

\section*{\textbf{Acknowledgements}}

We thank RUSA grant (Govt. of India) to Gauhati University for the High Performance Computing Facility in our computer Lab where this work was carried out..


\appendix
\section{$A_4$ product rules}
\label{appen1}
A discrete non-abelian group of even permutations of four objects called $A_4$ flavour symmetry is considered here to describe the flavour structure of fundamental particles. It is also the symmetry group of a tetrahedron. This group consist of  four irreducible representations: three one-dimensional and one three dimensional whose conventional notations are $\bf{1}, \bf{1'}, \bf{1''}$ and $\bf{3}$ respectively \cite{Ishimori:2012zz}. Their product rules are given as
$$ \bf{1} \otimes \bf{1} = \bf{1}$$
$$ \bf{1'}\otimes \bf{1'} = \bf{1''}$$
$$ \bf{1'} \otimes \bf{1''} = \bf{1} $$
$$ \bf{1''} \otimes \bf{1''} = \bf{1'}$$
$$ \bf{3} \otimes \bf{3} = \bf{1} \otimes \bf{1'} \otimes \bf{1''} \otimes \bf{3}_a \otimes \bf{3}_s $$
where $a$ and $s$ in the subscript refers to anti-symmetric and symmetric parts respectively. If $(a_1, b_1, c_1)$ and $(a_2, b_2, c_2)$ are two triplets, then their direct product can be decomposed into the direct sum as follows
$$ \bf{1} = a_1a_2+b_1c_2+c_1b_2$$
$$ \bf{1'} = c_1c_2+a_1b_2+b_1a_2$$
$$ \bf{1''} = b_1b_2+c_1a_2+a_1c_2$$
$$\bf{3}_s = (2a_1a_2-b_1c_2-c_1b_2, 2c_1c_2-a_1b_2-b_1a_2,$$
$$ \bf{2b_1b_2-a_1c_2-c_1a_2)}$$
$$ \bf{3}_a = (b_1c_2-c_1b_2, a_1b_2-b_1a_2, c_1a_2-a_1c_2)$$
\appendix


\end{document}